\documentclass[pre,aps,twocolumn,showpacs,superscriptaddress,floatfix]{revtex4-2}

\usepackage{mathtools}
\usepackage[usenames,dvipsnames]{xcolor}
\usepackage{amsmath}
\usepackage{amssymb,mathrsfs}
\usepackage{graphicx}
\usepackage[colorlinks=true]{hyperref}
\usepackage{soul}
\usepackage{xcolor}

\usepackage{subcaption}

\newcommand{\prt}{\partial}

\newcommand{\ga}{\gamma}
\newcommand{\om}{\omega}
\newcommand{\ka}{\kappa}

\newcommand{\ou}{\overline{u}}

\begin{document}

\title{Propagation of generalized Korteweg-de Vries solitons along large-scale waves}

\author{A.~M.~Kamchatnov}
\affiliation{Institute of Spectroscopy, Russian Academy of Sciences, Troitsk, Moscow, 108840, Russia}
\affiliation{Moscow Institute of Physics and Technology, Institutsky lane 9, Dolgoprudny, Moscow region, 141700, Russia}
\affiliation{Skolkovo Institute of Science and Technology, Skolkovo, Moscow, 143026, Russia}

\author{D.~V.~Shaykin}
\affiliation{Institute of Spectroscopy, Russian Academy of Sciences, Troitsk, Moscow, 108840, Russia}
\affiliation{Moscow Institute of Physics and Technology, Institutsky lane 9, Dolgoprudny, Moscow region, 141700, Russia}
\affiliation{Skolkovo Institute of Science and Technology, Skolkovo, Moscow, 143026, Russia}

\begin{abstract}
We consider propagation of solitons along large-scale background waves in the generalized Korteweg-de Vries (gKdV)
equation theory when the width of the soliton is mach smaller than the characteristic size of the
background wave. Due to this difference in scales, the soliton's motion does not affect the dispersionless
evolution of the background wave. We obtained the Hamilton equations for soliton's motion and derived simple
relationships which express the soliton's velocity in terms of a local value of the background wave. Solitons'
paths obtained by integration of these relationships agree very well with the exact numerical solutions of
the gKdV equation.
\end{abstract}

\pacs{05.45.Yv, 47.35.Fg}

% 05.45.Yv Solitons
% 42.65.Tg Optical solitons; nonlinear guided waves
% 47.35.Fg Solitary waves

\maketitle

\section{Introduction}

Perturbation theory for solitons has a long history and a number of publications is devoted to different
approaches to it which span from simple variational estimates to rigorous mathematical investigations based
on the inverse scattering transform method (see, e.g., review articles \cite{malomed-02,km-89} and references
therein). In spite of that, there still exist some specific situations where the developed so far methods
are either insufficient of too complicated for practical use and simpler approaches are needed. One such a
situation refers to propagation of solitons along a large-scale background wave $u=\ou(x,t)$, where
$\ou(x,t)$ obeys in the simplest case of unidirectional propagation to the Hopf-like equation
\begin{equation}\label{eq1}
  \ou_t+V_0(\ou)\ou_x=0.
\end{equation}
If we denote a characteristic width of a soliton as $\sim\ka^{-1}$, then it is assumed that $\ou(x,t)$
changes considerably at distances about $l$ much greater than $\sim\ka^{-1}$, so that $(\ka l)^{-1}\ll1$ is a
small parameter of the theory. At the same time, Eq.~(\ref{eq1}) is just a dispersionless approximation
of the nonlinear wave equation under consideration for unidirectional wave propagation. It is supposed that
soliton's propagation does not influence on evolution of the background wave, so this equation
does not contain any perturbation terms. This scheme corresponds to the generally accepted qualitative
picture according to which the soliton's propagation through a non-uniform and varying with time background
can be treated as motion of a classical particle under action of external time-dependent field.
Consequently, the first task is to derive equations for soliton's motion along the evolving background
wave $u=\ou(x,t)$. In fact, this problem was solved for Korteweg-de Vries (KdV) solitons in
Refs.~\cite{mt-79,mo-81} (see also \cite{mo-01,molotkov-05}), but this rigorous approach was quite
involved mathematically and was not, apparently, widely used in physical literature. Recently this
problem was reconsidered in Ref.~\cite{acehl-22} for propagation of KdV solitons along rarefaction waves
by different methods including the Whitham theory of modulations, and this approach was extended to the
problem of propagation of KdV solitons along dispersive shock waves (DSWs).

Application of the Whitham modulation theory to this type of problems seems very natural since propagation
of the soliton edge of a DSW reduces exactly to the motion of the leading soliton along the background
dispersionless wave. For example, this approach easily reproduces equations of motion \cite{ba-2000} in
Bose-Einstein condensate in case of absence of external perturbations (see Ref.~\cite{ik-22}). In this
paper we combine some ideas developed earlier in the Whitham modulation theory with elementary results
of the perturbation
theory and reproduce very simply the Hamilton equations of Refs.~\cite{mt-79,mo-81} for soliton's
motion. These equations can be integrated to give a useful relationship
\begin{equation}\label{eq2}
  \ka=\ka(\ou)
\end{equation}
between the soliton's inverse half-width $\ka$ and the background wave amplitude $\ou$. Since soliton's
velocity $V$ can be expressed in terms of the dispersion relation $\om=\om(k,\ou)$ for linear waves
with wave number $k$ propagating along the constant background $\ou$ by the Stokes formula \cite{stokes}
(it was published first in the early edition of Lamb's ``Hydrodynamics''\cite{lamb}, section 252;
see also \cite{kamch-20} and references therein)
\begin{equation}\label{eq3}
  V=\frac{\om(i\ka,\ou)}{i\ka},
\end{equation}
then substitution of Eq.~(\ref{eq2}) gives the equation
\begin{equation}\label{eq4}
  \frac{dx}{dt}=V(\ou(x,t))
\end{equation}
for soliton's path $x=x(t)$ which can be easily integrated.

The relationship (\ref{eq2}) can be treated as an analytical continuation of the relationship
between the carrier wave number $k$ of a short-wavelength wave packet and the background
amplitude $\ou$ which follows from the Hamilton theory of propagation of such packets \cite{ks-21}
as well as from the Whitham theory for propagation of small-amplitude edges of dispersive shock
waves \cite{el-05}. This observation allows us to extend the theory to the generalized KdV
equation and this is the main task of the present article. Our analytical results are confirmed
by exact numerical solutions of particular problems of propagation of solitons along large-scale
background waves.

\section{Motion of KdV soliton along a background wave}

The KdV equation
\begin{equation}\label{eq5}
u_t + 6uu_x+u_{xxx} = 0
\end{equation}
describes evolution of the whole wave structure
\begin{equation}\label{eq6}
  u(x,t)=\ou(x,t)+u_s(x,t)
\end{equation}
which consists of the background large-scale wave $\ou(x,t)$ obeying in our approximation
to the Hopf equation
\begin{equation}\label{eq7}
  \ou_t+6\ou\,\ou_x=0
\end{equation}
and the soliton
\begin{equation}\label{eq8}
  u_s(x,t)=\frac{\ka^2}{2}\cdot\frac1{\cosh^2[\ka(x-x_s(t))/2]},
\end{equation}
where $x_s(t)$ denotes the instant position of the soliton and $\ka=\ka(t)$ is its
time-dependent inverse half-width.

Our first task is to derive the Hamilton equations for the particle-like
motion of solitons. To this end, we notice that the dispersion relation for linear waves
propagating along constant background $u=\ou=\mathrm{const}$ reads
\begin{equation}\label{eq9}
  \om(k,\ou)=6\ou k-k^3.
\end{equation}
Then, according to the Stokes rule (\ref{eq3}), the soliton's velocity is given by
\begin{equation}\label{eq10}
  \frac{dx}{dt}=\frac{\om(i\ka)}{i\ka}= 6\ou(x,t)+\ka^2=\frac{\prt H}{\prt p},
\end{equation}
where we assume that the background $\ou=\ou(x,t)$ has the value corresponding to the
position $x$ of the soliton at the instant $t$. We expressed this velocity in Eq.~(\ref{eq10})
in Hamiltonian form, where the Hamiltonian $H=H(x,p)$ and the canonical momentum $p$ are to be
determined. The inverse half-width $\ka$ must be some function of $p$. We write this dependence
in the form $\ka^2=f(p)$, so integration of Eq.~(\ref{eq10}) gives
\begin{equation}\label{eq11}
  H=6\ou p+\int f(p)dp.
\end{equation}

To find $f(p)$, we need one more equation and, following Ref.~\cite{acehl-22}, we get it from the
first non-trivial conservation law for solitons. Substitution of Eq.~(\ref{eq6}) into (\ref{eq5})
gives
\begin{equation}\label{eq12}
  u_{s,t}+6(\ou u_s)_x+6u_su_{s,x}+u_{s,xxx}=-F[\ou(x,t)],
\end{equation}
where
\begin{equation}\label{eq13}
  F[\ou(x,t)]=\ou_t+6\ou\,\ou_x+\ou_{xxx}.
\end{equation}
We assume that $\ou(x,t)$ is a smooth solution of the Hopf equation (\ref{eq7}),
so the dispersion term in Eq.~(\ref{eq13}) can be neglected and we can take $F=0$. Then an
easy calculation with the use of several integrations by parts yields
\begin{equation}\label{eq14}
\begin{split}
  &\frac{d}{dt}\int_{-\infty}^{\infty}u_s^2dx=-12\int_{-\infty}^{\infty}u_s(\ou u_s)_xdx\\
  &=-6\int_{-\infty}^{\infty}\ou_xu_s^2dx\approx-6\ou_x(x,t)\int_{-\infty}^{\infty}u_s^2dx,
  \end{split}
\end{equation}
where we assume that since the distribution (\ref{eq8}) has a form of a narrow peak, the smooth
function $\ou_x$ can be replaced with good enough accuracy by its value at the soliton's position
$x$ and the moment $t$. With the use of Eq.~(\ref{eq8}) we find at once that
\begin{equation}\label{eq15}
  \int_{-\infty}^{\infty}u_s^2dx=\frac23\ka^3,
\end{equation}
so Eq.~(\ref{eq14}) transforms to the needed equation
\begin{equation}\label{eq16}
  \frac{d\ka^2}{dt}=-4\ou_x\ka^2.
\end{equation}

Now, substitution of Eqs.~(\ref{eq11}) and (\ref{eq16}) into the Hamilton equation
$$
\frac{dp}{dt}=-\frac{\prt H}{\prt x}
$$
gives with account of $p=p(f)=p(\ka^2)$
$$
\frac{dp}{df}\cdot\frac{d\ka^2}{dt}=-\frac{dp}{df}\cdot4\ou_xf=-\frac{\prt H}{\prt x}=-6\ou_xp,
$$
and, consequently, $2dp/p=3df/f$, so $p=f^{3/2}$ or $f=p^{2/3}$, where we have chosen the
integration constant equal to unity for simplicity of the notation.
At last, substitution of the function $f=p^{2/3}$ into Eq.~(\ref{eq11}) yields the Hamiltonian
in the form
\begin{equation}\label{eq17}
  H=6\ou(x,t)p+\frac35p^{5/3}.
\end{equation}
Soliton moves along the background wave $\ou=\ou(x,t)$ according to the Hamilton equations
\begin{equation}\label{eq18}
  \begin{split}
  &\frac{dx}{dt}=\frac{\prt H}{\prt p}=6\ou(x,t)+p^{2/3},\\
  &\frac{dp}{dt}=-\frac{\prt H}{\prt x}=-6\ou_x(x,t)p.
  \end{split}
\end{equation}
Equations (\ref{eq17}), (\ref{eq18}) coincide up to the notation with the equations obtained in
Refs.~\cite{mt-79,mo-81,mo-01} by a different method.

It is instructive to compare the above theory with the theory of Ref.~\cite{zf-71} where it was
shown that the KdV equation is a completely integrable Hamiltonian system in framework of the inverse
scattering transform method discovered in Ref.~\cite{ggkm-67}. The authors of Ref.~\cite{zf-71}
showed that the dynamics of any KdV wave reduces
to the Hamiltonian dynamics of the ``scattering data'' for the associated Schr\"{o}dinger
spectral problem, where solitons correspond to the discrete part of the spectrum and the ``background''
to its continuous part. In this exact theory, one does not separate characteristic sizes
of ``narrow'' solitons and ``wide'' background, so contributions of the background to the momentum
and energy are given by quite complicated formulas relating them with the scattering amplitude.
In our approximate approach the background dynamics is described by a simple dispersionless
equation (\ref{eq7}). At the same time, the both approaches lead to actually the same Hamiltonian
dynamics for a single soliton propagating along a zero background. In particular, according
to Refs.~\cite{zf-71,nmpz-80} the expressions for the momentum and energy are given in terms of
the inverse half-width $\kappa$ by the formulas
\begin{equation}\nonumber %\label{eq23b}
  \widetilde{p}=\frac13\ka^3,\quad \widetilde{H}=\frac15\ka^5,\quad\text{so that}\quad
  \widetilde{H}=\frac{3^{1/3}}{5}p^{5/3},
\end{equation}
which can be transformed to our formulas
\begin{equation}\nonumber %\label{eq23c}
  p=\ka^3,\quad H=\frac35\ka^5,\quad\text{so that}\quad H=\frac35p^{5/3},
\end{equation}
by multiplication of $\widetilde{p}$ and $\widetilde{H}$ by the constant factor 3 (obviously, the
Hamilton equations are invariant with respect to this transformation). Then addition of the term
$6\ou p$ to this Hamiltonian can be interpreted as the Galileo transform of the energy due to motion
of the background reference frame with velocity $6\ou$.

We have arrived at the Hamiltonian system (\ref{eq18}) which remains Hamiltonian when $\ou(x,t)$
evolves according to the hydrodynamic equation (\ref{eq7}). This means that the Poincar\'e-Cartan
integral invariant \cite{poincare,cartan}
\begin{equation}\label{eq19}
  I(C)=\oint_C(p\delta x-H\delta t)
\end{equation}
is preserved by the hydrodynamic flow (\ref{eq7}). As was shown in Ref.~\cite{kamch-23}, this
implies that the momentum $p$ of the moving soliton depends on $x$ and $t$ only via
the local value of the background variable $\ou(x,t)$. To find this dependence, we define first
a contour $C_0$ in the phase plane $(x,p)$ at $t=0$ and a tube of trajectories in the extended
space $(x,p,t)$ stemming from the initial points of $C_0$ and generated by the flow
$dx/dt=V_0(\ou(x,t))$, where $\ou(x,t)$ obeys Eq.~(\ref{eq1}). Next, we form an arbitrary
contour $C$ around this tube by means of introduction of a coordinate $\mu$ along each
trajectory and assume that $\mu$ is related with $x$ and $t$ by the equations
$$
\frac{dx}{V_0(\ou)}=dt=\chi(t,x,p(\ou))d\mu,
$$
where $\chi$ is an arbitrary function, so that the contour $C$ depends on the choice of $\chi$
and on the value of $\mu$. At last, we demand that $I$ does not depend on $\mu$, that is
$dI/dt=0$, then an easy calculation yields the condition
$$
\oint_C\left[\left(\frac{\prt H}{\prt p}-V_0\right)\frac{dp}{d\ou}+\frac{\prt H}{\prt\ou}\right]
\chi \left(\frac{\prt\ou}{\prt t}\delta t+\frac{\prt\ou}{\prt x}\delta x\right)=0.
$$
Consequently, in view of arbitrariness of the function $\chi$, we arrive at the equation
for the function $p=p(\ou)$,
\begin{equation}\label{eq20}
  \frac{dp}{d\ou}=\frac{\prt H/\prt\ou}{V_0(\ou)-\prt H/\prt p}.
\end{equation}
In our case with $V_0=6\ou$ and $H$ defined by Eq.~(\ref{eq17}), this equation takes the form
$p^{-1/3}dp=-6d\ou$, so we get
\begin{equation}\label{eq21}
  p^{2/3}=-4\ou(x,t)+q,
\end{equation}
where $q$ is an integration constant. According to our definitions above, $p^{2/3}=f=\ka^2$,
so this relation can be written as
\begin{equation}\label{eq22}
  \ka^2=-4\ou(x,t)+q.
\end{equation}
At last, substitution of Eq.~(\ref{eq21}) into the first Hamilton equation (\ref{eq18})
gives a very simple equation for the soliton's path
\begin{equation}\label{eq23}
  \frac{dx}{dt}=2\ou(x,t)+q,
\end{equation}
where it is assumed that $\ou(x,t)$ is a known solution of Eq.~(\ref{eq7}) and $q$ is
determined by the initial soliton's velocity. It is worth noticing that Eq.~(\ref{eq20})
transformed from $p$ to the variable $\ka=p^{1/3}$ coincides with the equation introduced
by G.~A.~El in Ref.~\cite{el-05} for description of motion of the soliton edge of dispersive
shock waves generated from evolution of step-like initial discontinuities.

Eq.~(\ref{eq22}) can be obtained in a simpler way with help of the Stokes reasoning \cite{stokes,lamb}
based on observation that the exponentially small soliton tails $u_s\propto\exp[\pm\ka(x-V_st)]$
and the small-amplitude harmonic waves $\propto\exp[i(kx-\om t)]$ obey the same linearized equations,
so the transformation $k\mapsto i\ka$ converts the phase velocity $V(k)=\om(k)/k$ into the soliton's
velocity (\ref{eq3}). From this point of view, the formula (\ref{eq3}) for the soliton velocity
is just an analytical continuation of the formula $V(k)=\om(k)/k$ for the phase velocity from the real
$k$-axis to its imaginary axis in the complex $k$-plane. This
idea of analytical continuation can be applied to some other expressions obtained for motion of
high-frequency wave packets converting them to relations between the soliton's parameters (see
Ref.~\cite{kamch-23}). For example, motion of a localized wave packet is described by its
coordinate $x=x(t)$ and carrier wave number $k=k(t)$ which obey the Hamilton equations
\begin{equation}\label{eq25}
  \frac{dx}{dt}=\frac{\prt\om}{\prt k},\qquad \frac{dk}{dt}=-\frac{\prt\om}{\prt x}.
\end{equation}
Again, when such a packet propagates along a background wave which evolves according to Eq.~(\ref{eq7}),
the wave number $k$ is a function of $\ou$ determined by the equation \cite{el-05,ks-21}
\begin{equation}\label{eq26}
  \frac{dk}{d\ou}=\frac{\prt \om/\prt\ou}{V_0(\ou)-\prt \om/\prt k},
\end{equation}
which in the KdV equation case (\ref{eq7}) and (\ref{eq9}) gives at once
\begin{equation}\label{eq27}
  k^2=4\ou-q,
\end{equation}
where $q$ is an integration constant. Analytical continuation of this formula to soliton's region
according to the Stokes rule $k\mapsto i\ka$ reproduces Eq.~(\ref{eq22}). The advantage of this
method, based on the well-established asymptotic theory of propagation of high-frequency wave packets,
is that it does not need derivation of Eqs.~(\ref{eq18}) or (\ref{eq20}). At last we notice that
formulas (\ref{eq22}) and (\ref{eq27}) also follow from the Whitham modulation equations
\cite{whitham-65,whitham} at their soliton and small-amplitude edges, correspondingly.

\begin{figure}[t]
\begin{center}
	\begin{subfigure}{3.7cm}	
	\caption{}
	\includegraphics[width = 3.7cm,height = 3.7cm]{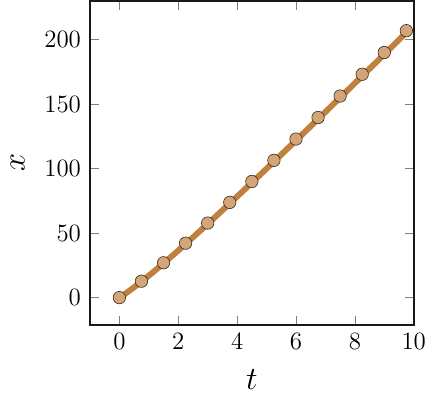}
	\end{subfigure}
\hspace{2mm}
	\begin{subfigure}{3.7cm}
	\caption{}
	\includegraphics[width = 3.7cm,height = 3.7cm]{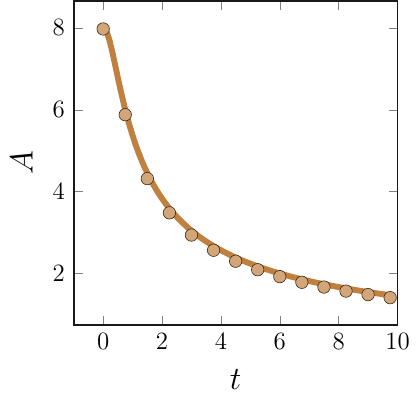}
	\end{subfigure}
\caption{(a) Soliton's path $x(t)$ obtained from solution of Eq.~(\ref{eq30}) (solid line) and
from exact numerical solution of the KdV equation (\ref{eq5}). (b) Change of the soliton's amplitude
$A$ during propagation of the soliton along the background wave (\ref{eq29}); solid line corresponds to
Eq.~(\ref{eq31}) and dots to the numerical solution of the KdV equation.
}
\label{fig1}
\end{center}
\end{figure}

Let us illustrate this theory by an example of propagation of a KdV soliton with the initial value
$\ka=4$, that is the initial amplitude $A=\ka^2/2=8$, which starts its motion at the point $x=0$ at
the moment $t=0$. The background wave has the initial profile
\begin{equation}\label{eq28}
  \ou(x,0)=(x/8)^2,\qquad x>0.
\end{equation}
Then we obtain from Eq.~(\ref{eq7}) the profile
\begin{equation}\label{eq29}
  \ou(x,t)=\frac{4}{9t^2}\left(\sqrt{1+\frac38xt}-1\right)^2,\qquad x>0,
\end{equation}
at any moment of time $t\geq0$. Since the initial velocity equals to $q=\ka^2=16$, Eq.~(\ref{eq23})
takes the form
\begin{equation}\label{eq30}
  \frac{dx}{dt}=\frac{8}{9t^2}\left(\sqrt{1+\frac38xt}-1\right)^2+16
\end{equation}
and it should be solved with the initial condition $x(0)=0$. The plot of this solution is shown in
Fig.~\ref{fig1}(a) by a solid line and dots correspond to soliton's positions obtained from numerical
solution of the whole KdV equation (\ref{eq5}) with the initial profile composed of the background wave
(\ref{eq28}) and the soliton (\ref{eq8}) located at $x=0$ with $\ka=4$. The solid line in
Fig.~\ref{fig1}(b) shows the analytical dependence (see Eq.~(\ref{eq22}))
\begin{equation}\label{eq31}
  A=\frac{\ka^2}{2}=8-2\ou(x(t),t)
\end{equation}
of the soliton's amplitude on time $t$ and dots correspond again to numerical values of the amplitude.
As we see, our approximate analytical theory agrees very well with the exact numerical solution.

\section{Generalized KdV equation}

Here we shall apply the above approach to the generalized KdV (gKdV) equation
\begin{equation}\label{eq32}
  u_t+V_0(u)u_x+u_{xxx}=0,
\end{equation}
where $V_0(u)$ is a monotonously growing function of $u$, $V_0(0)=0$. If we look for a soliton solution
in the form $u=u(\xi)$, $\xi=x-Vt$, when it propagates along a constant background $u\to\ou$ as
$|\xi |\to\infty$, then an easy calculation lead to the equation
\begin{equation}\label{eq33}
\begin{split}
  u_{\xi}^2=  G(u),\quad
  G(u)=  &V(u-\ou)^2-2\left[\Phi(u)-\Phi(\ou)\right]\\
  & +2\Phi'(\ou)(u-\ou),
  \end{split}
\end{equation}
where
\begin{equation}\label{eq34}
  \Phi(u)=\int_0^udu'\int_0^{u'}V_0(u^{\prime\prime})du^{\prime\prime}.
\end{equation}
The expression in the right-hand side of Eq.~(\ref{eq33}) has a double zero at $u=\ou$. One more zero
$u=u_m>\ou$ defines the soliton's amplitude by the equation
\begin{equation}\label{eq35}
  \frac12V(u_m-\ou)^2=\left[\Phi(u_m)-\Phi(\ou)\right]-\Phi'(\ou)(u_m-\ou),
\end{equation}
so that
\begin{equation}\label{eq36}
  A=u_m-\ou.
\end{equation}
The soliton's profile is determined in implicit form by the quadrature
\begin{equation}\label{eq37}
  \xi=\int_{u}^{u_m}\frac{du}{\sqrt{G(u)}}.
\end{equation}
We assume that this solution is stable. For example, in case of
\begin{equation}\label{eq38}
  V_0(u)=6u^{\ga}
\end{equation}
the soliton solution is stable, if $0<\ga<4$ (see Ref.~\cite{kuznetsov-84}).
In a particular case of modified KdV (mKdV) equation with $\ga=2$ the soliton solution can
be written in explicit form
\begin{equation}\label{eq39}
  u(\xi)=\ou+\frac{V-6\ou^2}{\sqrt{V-2\ou^2}\cosh(\sqrt{V-6\ou^2}\xi)+2\ou}.
\end{equation}
Consequently, in this case the inverse half-width $\kappa$ is related with the soliton's
velocity $V$ by the formula
\begin{equation}\label{eq40}
  \kappa^2=V-6\ou^2
\end{equation}
and the amplitude is equal to
\begin{equation}\label{eq40b}
  A=\frac{V-6\ou^2}{\sqrt{V-2\ou^2}+2\ou}.
\end{equation}

Now we turn to the problem of propagation of solitons along smooth large scale background waves.
In case of the gKdV equation (\ref{eq32}) we get the dispersion relation for harmonic waves
\begin{equation}\label{eq41}
  \om(k,\ou)=V_0(\ou)k-k^3.
\end{equation}
It is assumed that the wavelength is much smaller than a characteristic size of the
background wave $\ou$ whose evolution is described by Eq.~(\ref{eq1}). Now Eq.~(\ref{eq26})
gives (see \cite{el-05,kamch-20})
\begin{equation}\label{eq42}
  k^2=\frac23V_0(\ou)-q,
\end{equation}
$q$ is an integration constant. Analytical continuation of this formula to the soliton region
yields the expression for the soliton's inverse half-width
\begin{equation}\label{eq43}
  \kappa^2=q-\frac23V_0(\ou).
\end{equation}
Consequently, the Stokes rule (\ref{eq3}) gives the expression for the soliton's velocity
\begin{equation}\label{eq44}
   \frac{dx}{dt}=V=\frac{\om(i\ka,\ou)}{i\ka}=V_0(\ou)+\ka^2=\frac13V_0(\ou)+q,
\end{equation}
where the integration constant $q$ is determined by the initial conditions. Integration of this
equation with known law $\ou=\ou(x,t)$ of evolution of the background wave gives the path
$x=x(t)$ of the soliton. In case of the mKdV equation we get
\begin{equation}\label{eq44b}
  \frac{dx}{dt}=V=2\ou^2(x,t)+q
\end{equation}
and substitution of this expression for $V$ to Eq.~(\ref{eq40b}) gives the dependence of the
amplitude along the soliton's path,
\begin{equation}\label{eq44c}
  A(t)=\sqrt{q}-2\ou(x(t),t).
\end{equation}
Of course, this result can also be obtained directly from Eq.~(\ref{eq35}), which is applicable
to the general nonlinearity function $V_0(u)$,
with the use of Eqs.~(\ref{eq36}) and (\ref{eq44}).

The Hamilton equations for soliton's motion in case of the gKdV equation can be obtained without
much difficulty. We denote again $\ka^2=f(p)$ and integration of the equation
$$
\frac{dx}{dt}=\frac{\prt H}{\prt p}=V=V_0(\ou)+\ka^2=V_0(\ou)+f(p)
$$
gives
\begin{equation}\label{eq45}
  H=V_0(\ou)p+\int f(p)dp.
\end{equation}
Differentiation of Eq.~(\ref{eq43}) along the soliton's path yields
$$
\frac{d\ka^2}{dt}=-\frac23V_0'(\ou)(\ou_t+V\ou_x)=-\frac23V_0'(\ou)\ou_x\ka^2.
$$
Then the second Hamilton equation $dp/dt=-\prt H/\prt x$ gives
$$
\frac{dp}{df}\cdot\left(-\frac23V_0'(\ou)\ou_xf\right)=-V_0'(\ou)\ou_xp
$$
and, hence, $f(p)=p^{2/3}$, $p=\ka^3$. Thus, we obtain the Hamiltonian
\begin{equation}\label{eq46}
  H=V_0(\ou(x,t))p+\frac35p^{5/3}
\end{equation}
and the Hamilton equations
\begin{equation}\label{eq47}
  \frac{dp}{dt}=-V_0'(\ou)\ou_xp,\qquad \frac{dx}{dt}=V_0(\ou)+p^{2/3}.
\end{equation}

\begin{figure}[t]
\begin{center}
	\includegraphics[width = 7cm,height = 7cm]{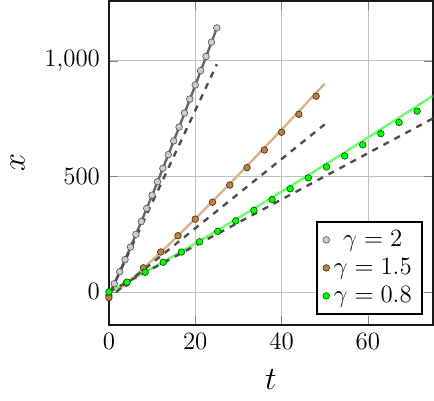}
\caption{Paths $x(t)$ of the solitons propagating along the background waves \eqref{eq49} for
different data sets ($\ga,a,v_0,x_0$): gray ($2,0.25,40,-20$), brown ($1.5,0.1,15,-25$) and green ($0.8,0.005,10,0$).
The circles correspond to the numerical solution of Eq.\eqref{eq32}, the solid lines to Eq.~\eqref{eq51},
and the dashed lines correspond to free motion of the soliton along a zero background.}
\label{fig2}
\end{center}
\end{figure}

Let us apply the developed theory to the problem of propagation of solitons in case of
the nonlinearity function (\ref{eq38}) and for the initial background wave distribution
\begin{equation}\label{eq48}
  \ou_0(x)=\left\{
  \begin{array}{ll}
  a(x/5)^{1/\ga}, & x>0,\\
  0, & x<0.
  \end{array}
  \right.
\end{equation}
Then the solution of the Hopf equation (\ref{eq1}) reads
\begin{equation}\label{eq49}
  \ou(x,t)=a\left(\frac{x}{t/\tau+5}\right)^{1/\ga},\quad \tau=\frac1{6a^{\ga}},
\end{equation}and $V_0(\ou)$ is given by
\begin{equation}\label{eq50}
  V_0(\ou(x,t))=\frac{x}{t+5\tau}.
\end{equation}
Let the soliton start its motion at the point $x=0$ at the moment $t=t_0>0$ with the
initial velocity $v_0$. Then Eq.~(\ref{eq44}) takes the form
\begin{equation}\label{eq51}
  \frac{dx}{dt}=\frac13\frac{x}{t+5\tau}+v_0
\end{equation}
and it can be easily solved to give
\begin{equation}\label{eq52}
  x(t)=\frac32v_0(t+5\tau)\left\{1-\left(\frac{t_0+5\tau}{t+5\tau}\right)^{2/3}\right\}.
\end{equation}
The soliton's amplitude along the path can be found from Eqs.~(\ref{eq35}), (\ref{eq36}) with
\begin{equation}\label{eq53}
  \Phi(\ou)=\frac{6}{(\ga+1)(\ga+2)}\ou^{\ga+2}
\end{equation}
and $V$ is defined by Eq.~(\ref{eq51}). These analytical predictions are compared with numerical
solutions of Eq.~(\ref{eq32}) in Fig.~\ref{fig2} and very good agreement is observed.

\section{Conclusion}

We showed that the KdV soliton dynamics along a large-scale background wave can be reduced to
Hamilton equations with the use of elementary perturbation theory argumentation. Preservation
of the Hamiltonian structure by the dispersionless flow leads to a simple relationship
between the inverse half-width of a moving soliton and a local value of the background wave.
This relationship can be interpreted as an analytical continuation of the relationship between the
carrier wave number of a wave packet propagating along a large-scale background wave which
follows from the well-known optical-mechanical analogy where the packet's dynamics is also
treated by the Hamilton methods. This type of reasoning first introduced by Stokes allows one
to extend the theory to the generalized KdV equation case and the analytical results are
confirmed by comparison with exact numerical solutions.

We believe that our approach based on preservation of Hamiltonian dynamics of both
high-frequency wave packets and narrow solitons by dispersionless hydrodynamic flow
can be applied to other problems of soliton dynamics.

\begin{acknowledgments}

This research is funded by the research project FFUU-2021-0003 of the Institute of Spectroscopy
of the Russian Academy of Sciences (Section~II) and by the RSF grant number~19-72-30028 (Section~III).

\end{acknowledgments}

\end{document}